# Observation of non-Hermitian topological disclination states and charge fractionalization

Ruifeng Li[1,2,†], Rimi Banerjee[2,†], Subhaskar Mandal[3,†], Da Li[1,†], Yang Long[2], Tianchi Ma[1], Jianwei Liu[2], Gui-Geng Liu[5,6], Yidong Chong[2,4,*], Baile Zhang[2,4,*], and Er-Ping Li[1,*]

[1] *College of Information Science & Electronic Engineering, Zhejiang Key Laboratory of Intelligent Electromagnetic Control and Advanced Electronic Integration, Zhejiang University, Hangzhou, 310027, China.*
[2] *Division of Physics and Applied Physics, School of Physical and Mathematical Sciences, Nanyang Technological University, Singapore, Singapore.*
[3] *Department of Physics, Indian Institute of Technology Bombay, Mumbai 400076, India*
[4] *Centre for Disruptive Photonic Technologies, Nanyang Technological University, Singapore, Singapore.*
[5] *Research Center for Industries of the Future, Westlake University, Hangzhou, 310030, China.*
[6] *Department of Electronic and Information Engineering, School of Engineering, Westlake University; Hangzhou, 310030, China.*

[†]*These authors contributed equally to this work.*
[*]*e-mail: yidong@ntu.edu.sg; blzhang@ntu.edu.sg; liep@zju.edu.cn*

## Abstract

There has been significant interest in exploring topological disclination states, which effectively probe the band topology of the host material beyond the conventional bulk-edge correspondence. While most studies in this area have primarily focused on Hermitian systems, recent theoretical work predicts that non-Hermiticity can drive topological phase transitions and host topological disclination states associated with fractional charge. However, no experimental observations have been reported to date. Here, we report the first experimental observation of topological disclination states in electric circuits, induced solely by gain and loss. Through admittance matrix measurements and eigenstate analysis, we confirm their emergence and compute the corresponding fractional charge. Moreover, the disclination mode profile and localization effect can be directly visualized via monochromatic field excitation. Additionally, we demonstrate the emergence of degenerate zero-energy topological disclination states, devoid of fractional charge, in distinct non-Hermitian geometries. Our findings open the possibility of non-Hermiticity-induced fractional charges in two-dimensional non-Hermitian lattices, which may pave the way for advancements in active topological photonic devices.

# Introduction

Topological materials represent a frontier in condensed matter physics, wherein defects serve as critical probes of fundamental properties[1–6]. Disclinations[7–12], defects of rotation symmetry, offer unique insights into bulk topology through trapped fractional charges[13,14]. This phenomenon emerges from the interplay between real-space topology of defects and momentum-space topology of the bulk bands, establishing a distinctive 'topological bulk-defect correspondence'[13]. Traditional boundary measurements often yield incomplete information due to symmetry reduction at boundaries. In contrast, disclinations enable direct probing of bulk topology through trapped fractional charges[15]. These trapped fractional charges arise from the fundamental incompatibility between charge quantization and lattice symmetries, making them particularly valuable for topological crystalline insulators and higher-order topological states[16–18], where conventional edge probes cannot fully characterize the bulk charge patterns.

Recent experimental validations of these theoretical predictions have established disclinations as robust probes for investigating topological matter[7,9,13,14,17,19,20]. However, existing platforms for topological bulk-defect correspondence commonly rely on Hermiticity as a basic assumption. Recent advances in non-Hermitian (NH) physics[21,22], from band theory[23–26] to topological lasers[27–29], have demonstrated significant potential in both theoretical frameworks and practical applications. NH topological phases exhibit characteristics fundamentally distinct from their Hermitian counterparts[30–34], generating substantial research interest. In particular, it has been recently proposed and experimentally demonstrated that non-Hermiticity can enable topologically protected boundary states in one-dimensional lattices[35–37] and in two-dimensional lattices[38,39]. Since classical-wave metamaterials systems often include non-negligible non-Hermitian effects, this naturally raises the question of whether non-Hermiticity can generate topological states confined to internal boundaries formed by topological defects, while also giving rise to intriguing phenomena such as charge fractionalization. Although this phenomenon has been theoretically predicted using tight-binding models[40], the experimental realization of fractional charges and localized states at NH disclinations has remained unexplored.

In this paper, we present the first experimental observation of NH disclination modes arising from charge fractionalization in electric circuit lattices, providing a convenient platform for precisely controlling gain and loss. Our experimental platform features two distinct unit cell designs with different gain-loss patterns, corresponding to topologically nontrivial and trivial phases. The gain and loss on-site potentials are realized through negative resistors and conventional resistors, respectively. Through admittance matrix measurements and eigenstate analysis of two C5-symmetric NH lattices, we observe a singlet disclination mode and two doublet disclination modes in the topological configuration, while the trivial configuration exclusively exhibits a pure bandgap. Quantitative analysis of the measured local density of states (LDOS) reveals that bulk unit cells surrounding the disclination core in the topological phase contribute a fractional charge of 1/2 modulo 1, while those in the trivial phase exhibit integral charge contributions. Monochromatic field excitation enables direct visualization of the disclination mode profile and its localization effects. Furthermore, investigations of C4-symmetric NH lattices reveal twofold degenerate midgap disclination states, behaving much like a NH version of the disclination states reported by Deng et al.[41], which could be advantageous for their spectral isolation and enhanced confinement[42].

## Results

As shown in Fig. 1a, by removing a π/3 sector (where the Frank angle equals π/3) from a graphene lattice with a trivial central defect and implementing a cut-and-glue procedure, we create a C5-symmetric lattice containing a disclination at the core. After adding on-site gain and loss distributions, we initially examine NH lattices exhibiting topologically nontrivial and trivial phases, as illustrated in Fig. 1b and Fig. 1c, respectively. The lattice is governed by the Hamiltonian:

$$H = \sum_{n}\left(i\gamma_n a_n^\dagger a_n\right) + t\sum_{\langle nn'\rangle}\left(a_{n'}^\dagger a_n + \text{h.c.}\right), \tag{1}$$

where $t$ represents the real-valued nearest-neighbor hopping amplitude, and $\gamma_n \in \mathbb{R}^+$ denotes the gain/loss level of the imaginary on-site mass term, taking values of $\gamma$ (gain) or $-\gamma$ (loss). $a_n^\dagger$ and $a_n$ are the creation and annihilation operators at site $n$, and $\langle nn'\rangle$ indicates nearest neighbors. The unit cell of this lattice, delineated by green solid lines, encompasses 18 sites.

The experimental realization of this tight-binding model is implemented through the electric circuit platform[43–50] depicted in Fig. 1d and 1e, corresponding to the NH lattices with topological and trivial phases, respectively. Each circuit diagram features a C6-symmetric unit cell, delineated by solid lines, with detailed circuit configurations shown in the inset. The hopping term $t$ is physically realized through coupling capacitors $C_t$. Individual sites of the discrete lattice model are implemented as resonance nodes, each connected to ground through a paralleled inductor $L_0$ and a parallel capacitor $C_0$. To obtain the less lossy spectra, making the admittance easier to measure over the noise, gain nodes are realized using a negative impedance converter with current inversion (INIC) in parallel with $L_0C_0$ resonators. The detailed circuit configurations for both gain and loss sites are illustrated in Fig. 1f. The INIC, an active non-reciprocal circuit element, continuously introduces energy into the system. Without proper regulation, this energy accumulation would drive the operational amplifier into nonlinear saturation, compromising its functionality. To prevent exponential energy growth and maintain system stability, we precisely calibrate the resistor values $R_a$ and $R_0$ (shown in Fig. 1f) to ensure the INICs operate within their linear gain regime. The specific parameter values, determined through comprehensive simulations, are detailed in Methods.

Governed by Kirchhoff's laws, the circuit system's response is described by $I = JV$, where $J$ denotes the circuit Laplacian, while $I$ and $V$ represent vectors of input currents and node voltages, respectively. At a given frequency $\omega$, the circuit Laplacian $J$ for NH lattices in Fig. 1d and 1e can be expressed as

$$J(\omega) \cdot i = \sum_n i \left( i\omega(C_0 + 3C_t) + \frac{1}{i\omega L_0} + \frac{1}{R_n} \right) a_n^\dagger a_n + \omega C_t \sum_{\langle nn' \rangle} \left( a_{n'}^\dagger a_n + \text{h.c.} \right), \qquad (2)$$

where $R_n$ assumes values of $R_0$ for gain sites and $-R_0$ for loss sites, the appendant term approaches zero when the frequency satisfies $\omega_0 = 1/\sqrt{L_0(C_0 + 3C_t)}$. Under this condition, $J$ exhibits a form analogous to the tight-binding Hamiltonian $H$. Consequently, the complex eigenvalues $E$ in tight-binding model map directly to the eigen-admittance $j$ in the circuit Laplacian at the resonance frequency $\omega_0 = 2\pi f_0$. The circuit component values are detailed in the Methods section, having been preselected to ensure $\gamma/t=1.3$ at the resonance frequency $f_0=361.72$kHz.

To provide a concrete demonstration and anticipate our experimental findings, we first present the complex energy spectrum of the NH topological circuit in Fig. 2a and Fig. 2b. The theoretical spectrum (Fig. 2a) is derived through numerical evaluation of the circuit Laplacian $J(\omega_0)$ using Equation (2), yielding the circuit lattice's eigen-admittance $j$. The detailed calculation of the circuit Laplacian and eigen-admittance $j$ is shown in Supplementary Note 1. Furthermore, the experimental spectrum (Fig. 2b) is obtained through circuit Laplacian characterization using a vector network analyzer (VNA) (also see in Methods). The methods for identifying both the bulk gap (indicated by red vertical dashes) and the occupied states are described in Supplementary Note 2. Both calculation and experimental results show good agreement in revealing five disclination states and corner modes in the mid-gap region. Moreover, Fig. 2c illustrates experimental observations of five disclination states within the bulk gap, comprising one singlet mode (indicated by the blue dot in Fig. 2b) and two doublet modes (indicated by the yellow and red dots in Fig. 2b). The phase properties of disclination modes are detailed in Supplementary Note 3. The spatial voltage distributions of these five disclination states exhibit pronounced localization around the disclination core, demonstrating excellent concordance with theoretical predictions. Following similar numerical and experimental approaches, we analyze the complex energy spectra for the topologically-trivial circuit (illustrated in Fig. 1d) and present the results in Fig. 2d and Fig. 2e, respectively. These spectra demonstrate a pure bandgap devoid of mid-gap states, confirming the absence of disclination modes in the trivial phase (mode profiles near the bulk gap are given in Supplementary Note 4).

Next, we turn to the calculation of fractional charge of bulk unit cells near the disclination. In Hermitian lattices, a similar argument for charge fractionalization has been verified in experiments on microwave metamaterials[14], photonic crystals[13], and elastic plates[19]. Theoretical prediction suggests that NH disclination lattices with specific gain-loss distributions exhibit anomalous electron filling, resulting in fractional charges around disclinations[40]. Following this prediction and adopting the established methodology of analyzing LDOS in circuit lattices, we conduct both numerical calculations and experimental measurements on the C5-symmetric circuit in its topological phase. Following the canonical definition, the circuit disclination charge for each bulk unit cell around the disclination core can be determined as[40,51,52]:

$$Q_u = \sum_n \sum_l \left| \langle V_n^L | l \rangle \langle l | V_n^R \rangle \right| (\mathrm{mod}\, 1), \qquad (3)$$

where $\left| V_n^{R/L} \right\rangle$ represent the bi-orthogonal $n$-th right/left eigenvector of the full circuit Laplacian $J$, which are obtained numerically from the measured circuit Laplacian. $Q_u$ denotes the disclination charge for a bulk unit cell, with $l$ summed over 18 sites in each bulk unit cell plus 2 nearest sites in disclination core, and $m$ summed over all occupied states.

The $Q_u$ calculation details are illustrated in Supplementary Note 5. In our NH circuit lattice, the trapped disclination charge in each bulk unit cell clusters around 1/2 mod 1, yielding total disclination charges of 0.57 and 0.47 for calculation and experiment, respectively. For comparison, the respective spectral charge distributions of the circuit with trivial phase reveal charge values concentrating around 0 mod 1, with total disclination charges 0.11 and 0.10 for calculation and experiment, respectively. These comprehensive results demonstrate reasonable consistency between numerical calculations and experimental observations, illustrating an experimental observation of charge fractionalization in NH topological circuit lattices.

Having observed the disclination modes and charge fractionalization phenomena through circuit Laplacian experiments, we implement direct visualization of localization effects in the circuit lattice (also found in Methods). We apply an excitation scheme based on the phase characteristics of a disclination mode from one of the doublets to investigate the localization behavior. The core intensity is quantified by calculating the sum of squared voltage amplitudes across all five disclination cores. Fig. 3a illustrates the frequency-dependent core intensity profiles for various non-Hermiticity parameters $\gamma$, implemented through adjusting the parallel resistor values $R_0$ in the resonators. The observed shifts in resonant frequency across different $\gamma$ values demonstrate strong agreement with theoretical predictions. The experimental measurements, indicated by green dots, are obtained through linear superposition of individual excitations at the disclination cores with precisely controlled phase differences, utilizing a vector network analyzer (VNA) as shown in the inset (detailed voltage distributions are given in Supplementary Note 7). The experimental observation of the disclination mode shows great agreement with simulation for $\gamma$=1.3.

Due to the dispersive nature of circuit element admittance, observation of the circuit lattice's topological properties in the frequency domain requires the circuit Hamiltonian[53]. Fig. 3b and Fig. 3c illustrate the eigenstate spectra as a function of eigenfrequency for circuits in topological and trivial phases, respectively. These frequency-domain eigenstate spectra, transformed from the measurements presented in Fig. 2b and Fig. 2e, preserve the topological properties of the C5-symmetric lattices. For our experimental observations, we selected the frequency $f$=353.2 kHz corresponding to the first doublet disclination mode (indicated by horizontal red dashed lines) (the circuit Hamiltonian transformation is detailed in Supplementary Note 8). The steady-state voltage responses, measured via VNA at this selected frequency, are displayed in the insets, demonstrating remarkable concordance between numerical simulations and experimental measurements. Notably, the topological circuit exhibits strong localization effects, manifesting as clearly observable disclination mode profiles at the observation frequency, shown in Fig. 3b. In contrast, the trivial circuit configuration (Fig. 3c) displays no such localization, as the observation frequency lies within a spectral gap where mode excitation is suppressed.

We further expanded our investigation to include a C4-symmetric NH circuit configuration by removing a 2π/3 sector from the NH honeycomb lattice, as shown in Fig. 4a. Fig. 4a presents the fabricated C4-symmetric circuit, with the inset providing a detailed view of the unit cell structure. The eigenstate spectra, derived from both theoretical calculations (Fig. 4b) and experimental measurements (Fig. 4c), reveal distinct degenerate mid-gap states, serving as a non-Hermitian analogue to the results reported by Deng et al.[41]. The insets display the voltage distribution patterns of these disclination states, encompassing both amplitude and phase information. The corresponding complex energy spectra from theoretical calculations (Fig. 4d) and experimental measurements (Fig. 4e) demonstrate good agreement, clearly indicating the emergence of zero-energy topological disclination states. In contrast, the C4-symmetric NH circuit with an alternative gain/loss pattern (Fig. 4f) exhibits substantially different characteristics. The complex energy spectra obtained from calculations (Fig. 4g) and experiments (Fig. 4h) confirm the absence of disclination states. Details about C4-symmetric circuits can be found in Supplementary Note 9.

## Conclusion

To summarize, we experimentally demonstrate NH disclination modes and fractional charges through precise measurements of circuit Laplacians and eigenstates in NH electric circuit lattices. By extending the bulk-defect correspondence to the non-Hermitian regime, we present the first experimental observation of NH disclination modes generated by strategically patterned gain and loss distributions at lattice sites. As key manifestations of this correspondence, we observe both the charge fractionalization effect, characterized by a disclination charge of 1/2, and the mode localization effect around the disclination core, extending beyond the traditional Hermitian assumptions. Furthermore, we report the first observation of degenerate zero-energy topological disclination states in the NH regime. Notably, while our setup has been carefully designed to operate within the linear regime, it can also be adapted to function in the nonlinear regime. In the future, exploring this nonlinear regime, including the effects of gain saturation, could provide valuable insights and potentially contribute to the development of novel topological lasers.

# Method

**Circuit sample construction**

Circuit elements are selected and validated using the LTspice simulation software. The NH disclination circuits are constructed using pre-selected components with the following specifications: capacitors $C_t=C_0$=2.2nF (±1%), inductors $L_0$=22μH (±1%), and resistors $R_0$=154Ω (±5%), configured to achieve $γ/t$=1.3. The INIC is implemented using the LM6171 unity-gain stable operational amplifiers (Texas Instruments). To ensure operation within the linear region of the LM6171, we select $R_a$=300Ω and verify stability through simulation. At the ten disclination core sites, which feature reduced circuit connections, we employ $C_0$=4.4nF (±1%) capacitors to maintain the Laplacian form.

The printed circuit board (PCB) is designed with ground layers inserted between the source and signal layers to minimize electromagnetic coupling. Signal traces are implemented with a width of 0.508 mm to minimize parasitic inductance, while power supply traces are designed at 0.635 mm to enhance their operational stability. Components are spaced at minimum intervals of 1.5 mm to reduce inductive coupling interference. The C5-symmetric PCBs comprise 15 circuit unit cells and 10 disclination sites, totaling 280 sites across a 569 mm × 548 mm board area.

**Experimental measurements**

Admittance eigenvalues and eigenstates are characterized through scattering matrix (**S**) measurements using a Vector Network Analyzer (VNA, Keysight E5061B). The operational amplifiers are powered by a DC supply providing ±15 V. The impedance matrix **Z** is derived using the microwave network transformation: $\mathbf{Z} = Z_0(\mathbf{I} + \mathbf{S})(\mathbf{I} - \mathbf{S})^{-1}$, where $Z_0$ represents the VNA port impedance and **I** denotes the identity matrix. This impedance matrix **Z**, being the inverse of the circuit Laplacian $J(\omega_0)$, provides comprehensive information about the admittance eigenvalues and eigenstates. Direct circuit site excitation measurements are also performed using the VNA (Keysight E5061B), with subsequent vector superposition post-processing.

# Code availability

Circuit simulations are performed using LTspice (https://www.analog.com/en/design-center/design-tools-and-calculators/ltspicesimulator.html#).

## Acknowledgements

The work at Zhejiang University was sponsored by the National Natural Science Foundation of China (NNSFC) under Grants No. 62071424, 62201499 and 62027805.

## Author contributions

R.L., R.B., and S.M. initiated the project, with discussions from Y.C., B.Z., and E.L. R.L., R.B., S.M., G.-G.L., Y.L., and J.L. performed the theoretical calculations. R.L. conducted the circuit simulations, designed the experiments, and analyzed the data with assistance from R.B. and S.M. R.L., D.L., and T.M. carried out the measurements. All authors discussed the results and came to an agreement on the conclusions. R.L., R.B., S.M., Y.C., B.Z., and E.L. wrote the manuscript. Y.C., B.Z., and E.L. supervised the project.

## Competing interests

The authors declare no competing interests.

# Figures

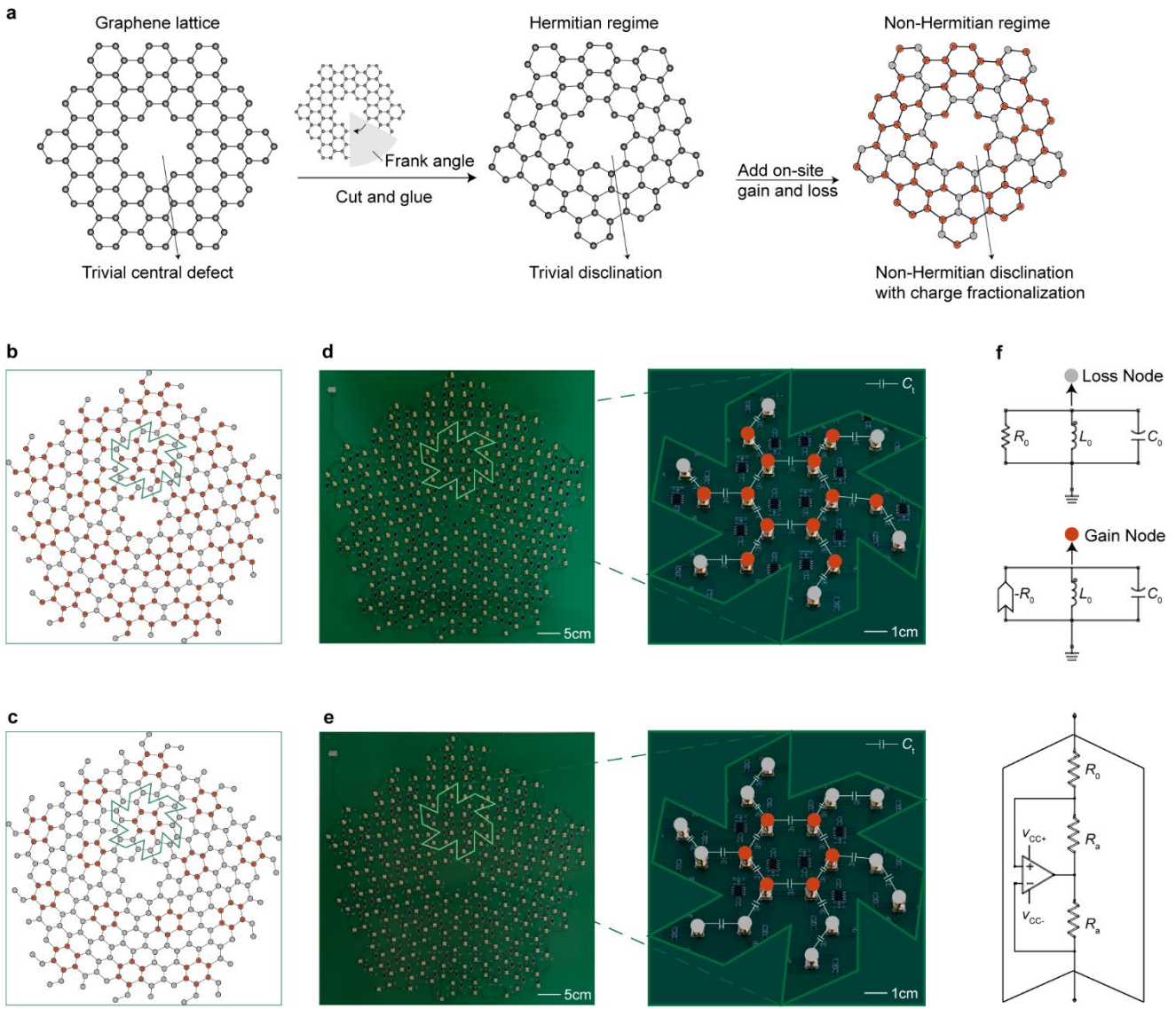

**Fig. 1 | The photograph shows the fabricated samples and corresponding NH lattices. a** Illustration of the formation of the non-Hermitian disclination lattice. **b, c** Schematic of NH lattices containing a disclination, with topological phase (**b**) and trivial phase (**c**). **d, e** Photographs of the fabricated electric circuits for C5-symmetric disclination lattices with a topological phase (**d**) and a trivial phase (**e**). The C6-symmetric solid line box outlines a unit cell and the inset presents its enlarged view. **f** The circuit representation of the on-site potential in NH lattices. On-site loss depends on resistors while on-site gain depends on negative resistors.

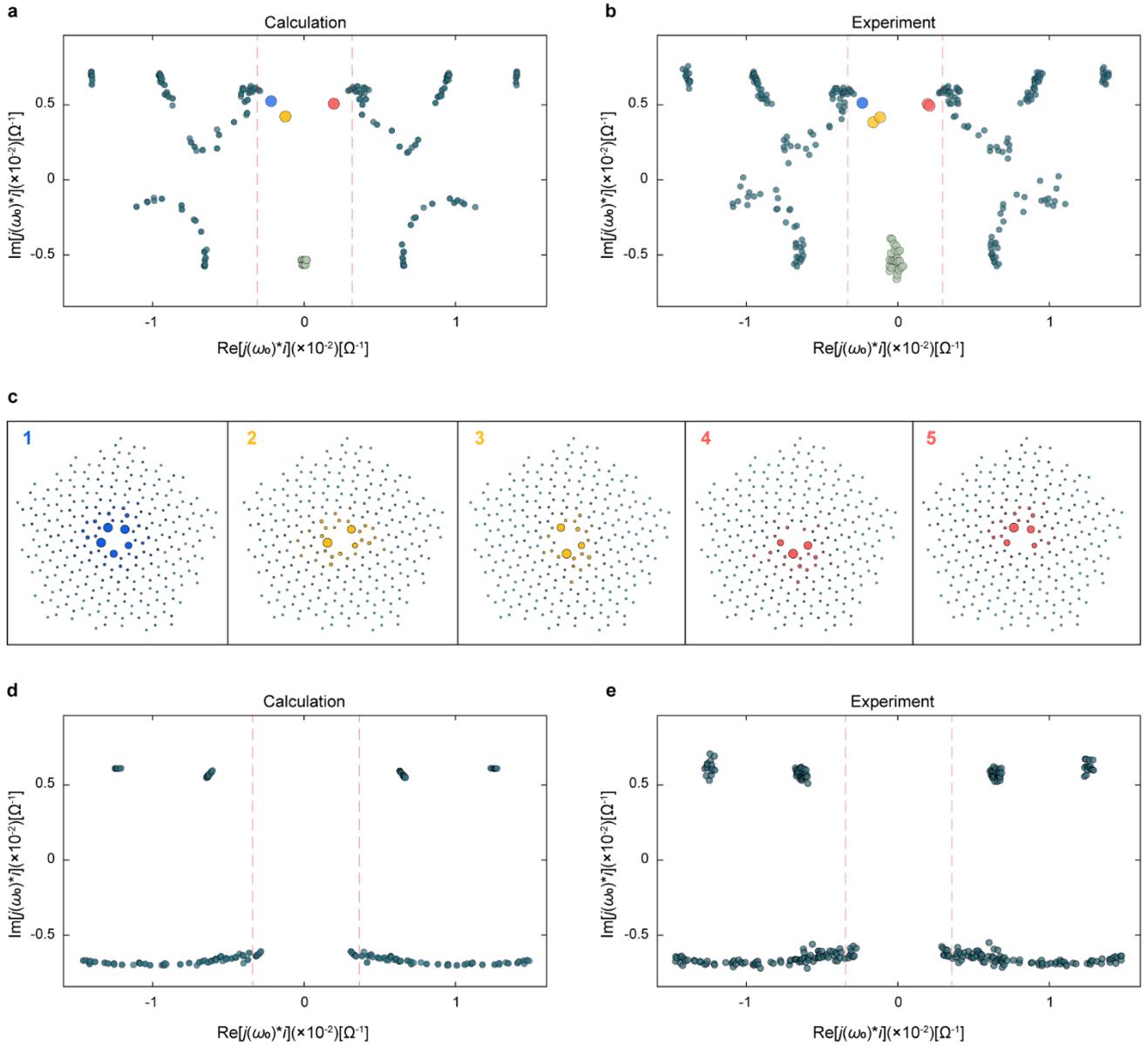

**Fig. 2 | Experimental observation of disclination modes in C5-symmetric NH electric circuit. a, b** Complex energy spectrum for the NH topological circuit derived by numerical calculation (**a**) and experimental measurements (**b**). The blue dot denotes the singlet mode, whereas the yellow and red dots correspond to two doublet modes. The light green clusters represent corner modes, while the dark green dots indicate edge modes and bulk states. **c** Measured voltage distributions at the resonance frequency for the five disclination states in (**b**). **d, e** Complex energy spectrum for the NH topologically-trivial circuit derived by numerical calculation (**d**) and experimental measurements (**e**), showing a pure bandgap.

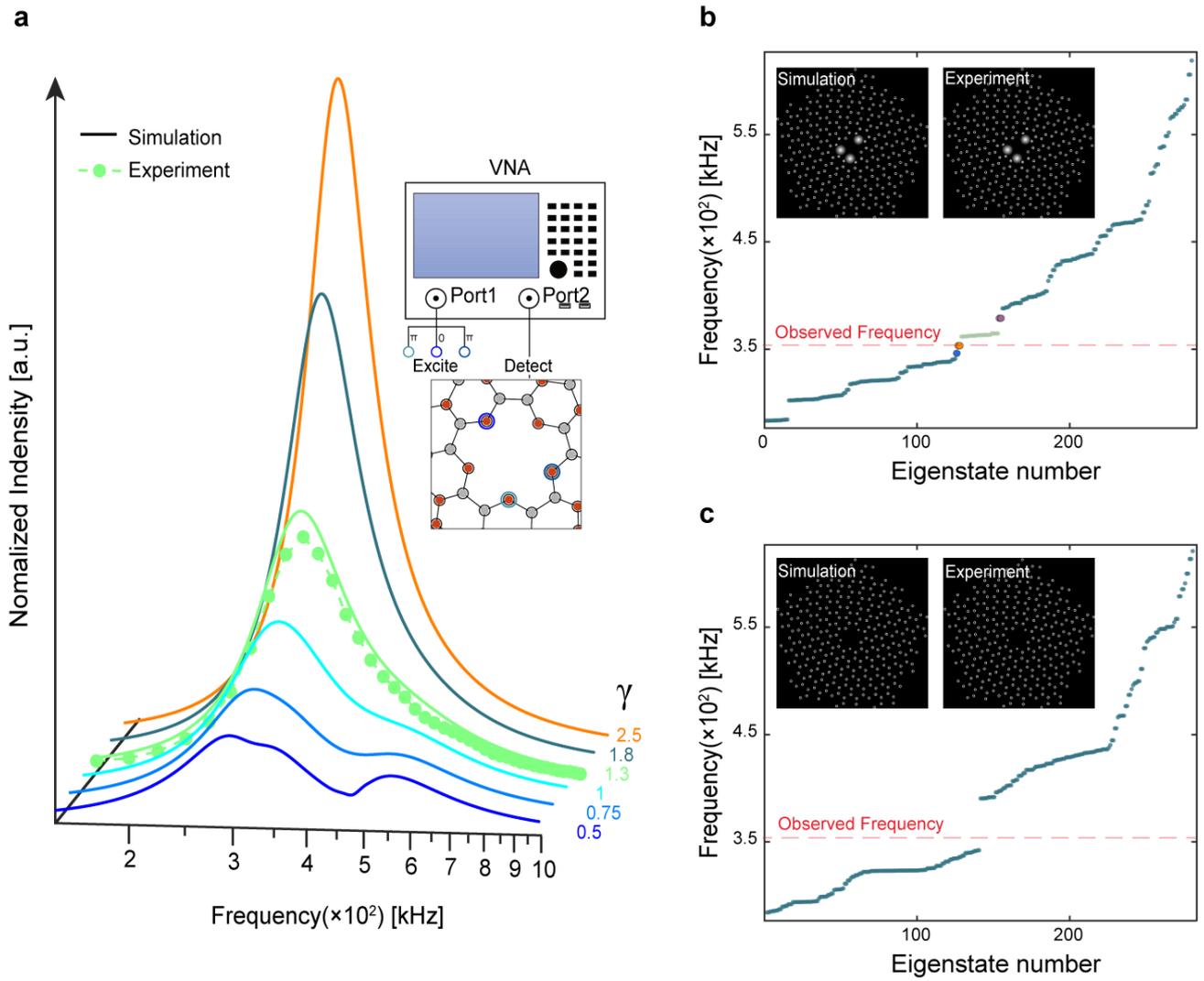

**Fig. 3 | Direct visualization of the disclination mode profile and localization effect**. **a** Frequency-dependent intensity distribution illustrating the disclination mode profile in a C5-symmetric circuit. Simulation results (solid lines) depict the behavior across varying gain/loss parameters $\gamma$ (0.5-2.5), while experimental measurements (green dashed line) at $\gamma = 1.3$ are obtained using the vector network analyzer (VNA) configuration shown in the inset. **b, c** The eigenstate spectra of the circuit are plotted as a function of resonant frequency, comparing the topological phase (b) and trivial phase (c). The insets show steady-state voltage responses at the observed frequency (indicated by red dashed lines), with both simulated and experimental results displayed.

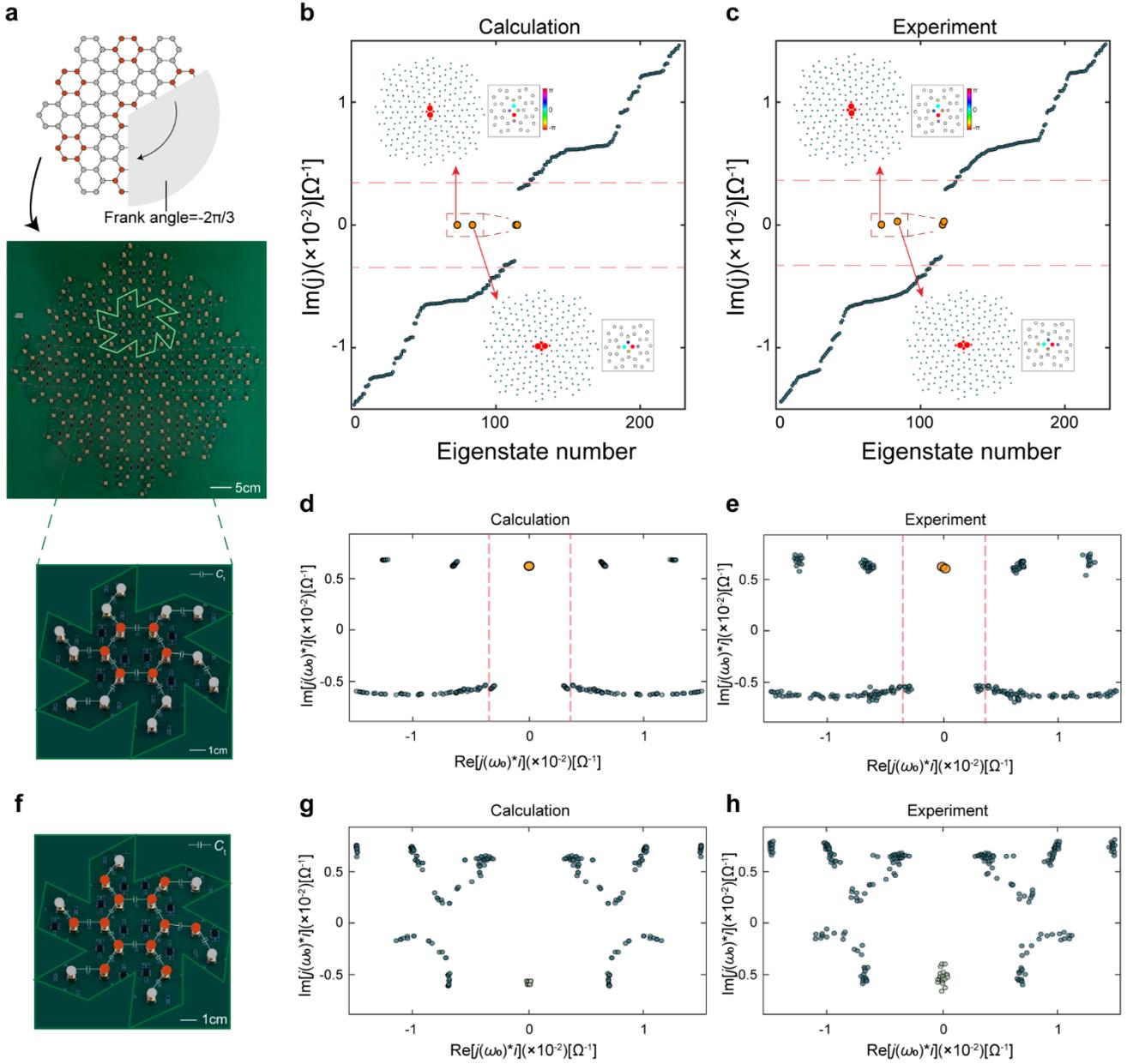

**Fig. 4 | Degenerate zero-energy topological disclination states in C4-symmetric NH circuit. a** Illustration of the C4-symmetric NH lattice formation and the photograph of the fabricated C4-symmetric NH circuit. The inset shows the enlarged view of the corresponding unit cell. **b, c** Eigenstates spectra of topological circuit obtained through calculation (**b**) and experiment (**c**). The inset depicts the voltage distributions of two mid-gap disclination states, including both amplitude and phase information. **d, e** Complex energy spectrum derived from calculation (**d**) and experiment (**e**). **f** The photograph of the fabricated C4-symmetric NH circuit unit cell with an alternative gain/loss configuration. **g, h** Complex energy spectrum obtained through calculation (**g**) and experiment (**h**).

16